\documentstyle{article}

\def\bichpr{\hoffset=-20truemm
\voffset=-20truemm
\textwidth=16 truecm
\textheight=24 truecm }

\bichpr

\begin{document}
\title{A three-fermion Salpeter equation.} 
\author{ J. Bijtebier\thanks{Senior Research Associate at the
 Fund for Scientific Research (Belgium).}\\
 Theoretische Natuurkunde, Vrije Universiteit Brussel,\\
 Pleinlaan 2, B1050 Brussel, Belgium.\\ Email: jbijtebi@vub.ac.be}
\maketitle
\begin{abstract}   \noindent
  We write a 3D equation for three fermions   by combining the three two-body potentials obtained by the
 reductions of the corresponding two-fermion Bethe-Salpeter equations to equivalent 3D equations, putting the spectator fermion
on the mass shell. In this way, the cluster-separated limits are still exact, and the   Lorentz invariance / cluster
separability requirement is automatically satisfied, provided no supplementary approximation, like the Born approximation, is
made. The use of positive free-energy projectors in the chosen reductions of the two-fermion Bethe-Salpeter  equations   prevents
continuum dissolution in our 3D three-fermion equation. The potentials are hermitian and depend only slowly on the total
three-fermion energy.  The one high-mass limits are approximately exact. \par
In view of a possible perturbation calculation, correcting the remaining discrepancies with the
three-fermion Bethe-Salpeter equation, we succeeded in deriving our 3D equation from an approximation of the three-fermion
Bethe-Salpeter equation, in which the three-body kernel is neglected and the two-body kernels approached by positive-energy
instantaneous expressions, with the spectator fermion on the mass shell. The neglected terms are transformed into corrections to
the 3D equation. A comparison is made with Gross' spectator model.
\end{abstract}
equations. \newline
Salpeter's equation. Breit's equation.\par  Relativistic bound states. Relativistic wave equations. \\\\
 \newpage
\tableofcontents

\section{Introduction}
In the treatment of the three-body problem, there is a large gap between nonrelativistic quantum mechanics (Schr\"odinger
equation) and relativistic quantum field theory (Bethe-Salpeter equation \cite{1,2}). Starting from the Schr\"odinger
equation one can of course replace the free part of the hamiltonian by its relativistic form. Starting from the Bethe-Salpeter
equation, one can try to eliminate the two relative time variables to get finally a Schr\"odinger equation with a compact
potential plus a lot of correction terms of various origins. Instead of the Schr\"odinger equation, one can use the Faddeev
equations. These equations can be derived from the Schr\"odinger equation, or, in a more general form, from the Bethe-Salpeter
equation. Faddeev equations for the transition operator give the various scattering and reaction matrix elements and the poles
(to be computed by basically nonperturbative methods) of this operator in the total energy give the spectrum of the
three-body bound states.\par
Schr\"odinger's equation is "relatively easy" to solve, but this zero-order approximation does not reflect
several important properties and symmetries of the studied physical system. These could be recovered, in principle,
by incorporating higher-order contributions (in general an infinity of them).\par
As an intermediate step between nonrelativistic quantum mechanics and quantum field theory we shall search for a 3D equation
built with the sum of the relativistic free hamiltonians plus two-body and perhaps also three-body interaction potentials. Such
equations are closely related to the systems of coupled Dirac or Klein-Gordon equations of constraint theory
\cite{22,23,24,18,26}, which are still 4D equations but exhibit a
simplified dependence in the relative times, which could be completely eliminated to get a single 3D equation.  Besides
containing relativistic free hamiltonians, we shall try to make our 3D equation satisfy the following
list of requirements:\par\hfill\break\indent
--- Correct nonrelativistic limit.\par
--- Lorentz invariance and cluster separability. It is always possible to render an equation Lorentz
invariant by working in the general rest frame (center of mass reference frame) and by building invariants with the
total 4-momentum vector, although the result may be unelegant and artificial. The Lorentz invariance
requirement becomes a tool when combined with the cluster separability requirement: when all mutual
interactions are "switched off", we must get a set of 3 free Dirac equations. This total separability can 
easily  be obtained by using as hamiltonian the sum of three free Dirac hamiltonians and interaction terms. The real difficulty
appears when only the interactions with fermion 3 (for example) are switched off. If we want a full cluster
separability, the resulting equation for the (12) cluster can not refer to the global center of mass frame anymore, as the
momentum of fermion 3 enters in the definition of this frame. \par
Lorentz invariance and cluster separability can be explicit or implicit (via rearrangements). The best known
example of implicit Lorentz invariance is a free Dirac equation solved with respect to the energy: it
becomes explicitly covariant by multiplication with the $\,\beta\,$ matrix. Other implicit Lorentz invariances are
not that trivial. For example, the 3D reductions of a Bethe-Salpeter equation are implicitly covariant, provided the series
generated by this reduction is not truncated.\par
 --- Hermiticity and total energy independence of the interaction terms. In the
two-body problem, nonhermitian interaction terms can be hermitian with respect to a modified scalar product, or
made hermitian via a rearrangement of the equation. In the three-body problem these rearrangements could be
more complicated. The hermiticity and the independence in the total energy are linked features, as one of
these is often achieved at the expense of the other one. Energy dependending interaction terms destroy some of
the advantages of the use of an hermitian hamiltonian, such as the mutual orthogonality of the solutions,
and leads to modify the usual perturbation calculation methods. The 3D potentials deduced from
field theory are generally energy dependent, at least in the higher-order terms. We shall require  hermiticity and
energy-independence (or slow energy  dependence) in the lowest order terms at least.\par 
--- Correct heavy mass limits. When the mass of one of the
fermions becomes infinite, its presence must be translated in the equations by a potential (Coulombian in QED) acting on the other
fermions. For the two-body problem the "one-body limit" in QED is indeed a Dirac-Coulomb equation or a rearrangement of
it (for example when a projection operator is introduced in order to avoid continuum dissolution - see
below). In the Dirac-Coulomb equation, the Coulomb potential is already given by the limit of the Born term, as the higher-order
crossed and ladder terms cancel mutually at the one-body limit. In contrast, the one-body limits of the rearrangements
contain contributions from all terms \cite{3,4,5,6,7} .\par
--- Perturbative approach. It must be possible to start with a manageable approximate equation and indefinitely improve the
approximation of the measurable quantities (with respect to the uncalculated predictions of the here assumed exact Bethe-Salpeter
equation) by adding higher-order contributions.\par
--- Solution of the continuum dissolution problem. In the relativistic equations for several
relativistic particles, the physical bound states are degenerate with a continuum of states
combining asymptotically free particles with opposite energy signs. This often neglected fact
forbids the building of normalizable solutions in the $\,N\!>\!2$-body problem (including the two-body plus
potential problem. In the pure two-body problem the mixing is prevented by the conservation of the
total momentum). The usual solution consists in including positive-energy projection operators into the
zero-order propagator \cite{8,9,10,11}. The modified equations must of course continue to satisfy the other
requirements, like the Lorentz invariance / cluster separability requirement.\par\hfill\break\indent
 Section 2 is devoted to the two-fermion problem, revisited in order to define the notations and present the
building blocks and cluster-separated limits of our future three-fermion equations.  In section 3 we write directly a cluster
separable 3D equation, by combining the relativistic free hamiltonians and the 3D potentials obtained in the reduction of the
two-fermion Bethe-Salpeter equations, putting in each 3D potential the spectator fermion on the mass shell. Switching off two
of the three mutual interaction potentials gives then a free Dirac equation for the spectator fermion plus a two-body equation,
equivalent to the two-fermion Bethe-Salpeter equation, for the two interacting fermions. This equivalence insures the Lorentz
invariance / cluster separability property.
Positive-energy projectors solve the continuum dissolution problem. Section 4 is devoted to the computation of the
heavy-mass limits of the three-fermion equation, which are compared with the two-fermion in an external potential equations.
In section 5, we show that our 3D equation can be obtained from the three-fermion Bethe-Salpeter equation, by neglecting the
three-body kernel and replacing the two-body kernels by positive-energy instantaneous approximations, equivalent at the
cluster-separated limits. The neglected terms are thus known at the Bethe-Salpeter equation level. A comparison is made with
Gross' spectator model \cite{12,13}. This section ends with an attempt of transforming the neglected terms in the Bethe-Salpeter
equation into correction terms to the 3D equation, in view of a possible perturbation calculation.   Section 6 is devoted to
conclusions.
   
\section{The two-fermion problem.}
\subsection{Notations.}
\noindent We shall write the Bethe-Salpeter equation for the bound states
 of two fermions \cite{1} as 
\begin{equation}\Phi = G_0 K \Phi,    \label{1}\end{equation}   where $\Phi$ is the Bethe-Salpeter  amplitude, function of the
positions $x_1,x_2$ or of the momenta 
$p_1,p_2$ of the fermions, according to the representation chosen. The operator $K$  is the Bethe-Salpeter kernel, given in a
non-local momentum representation by the sum of the irreducible two-fermion Feynman graphs. The operator $G_0$ is the free
propagator, given by the product
$G_{01}G_{02}$ of the two individual fermion propagators
\begin{equation} G_{0i} = {1 \over p_{i0}-h_i+i\epsilon h_i}\,\beta_i = {p_{i0}+h_i\over p_i^2-m_i^2+i\epsilon}\,\beta_i 
\label{2}\end{equation}  where the $h_i$ are the Dirac free hamiltonians
\begin{equation}h_i = \vec \alpha_i\, . \vec p_i + \beta_i\, m_i\qquad (i=1,2).  \label{3}\end{equation}  The Bethe-Salpeter
kernel $\,K \,$ should contain charge renormalization and vacuum polarization graphs, while the propagators $\, G_{0i}\,$ should
contain self-energy terms (which can be transferred to $\, K\,$ \cite{14,15} ]). In this work, we consider only the free fermion
propagators in   $\,G_{0i}\,$ and the "skeleton" graphs in $\,K$, and we hope that the inclusion of the various corrections would
not change our conclusions.\par\noindent  We shall define the total (or external, CM, global) and relative (or internal) variables:
\begin{equation}X = {1 \over 2} (x_1 + x_2)\ , \qquad P = p_1 + p_2\ ,  \label{4}\end{equation} 
\begin{equation}x = x_1 - x_2\ , \qquad p = {1 \over 2} (p_1 - p_2). \label{5}\end{equation}  and give a name to the
corresponding combinations of the free hamiltonians:
\begin{equation}S = h_1 + h_2\ , \quad s = {1 \over 2} (h_1 - h_2).\label{6}\end{equation}  We know that, at the no-interaction
limit, we shall have to get a pair of free Dirac equations:
\begin{equation} (p_{10}-h_1)\Psi=0, \qquad (p_{20}-h_2)\Psi=0,  \label{7}\end{equation}  where $\,\Psi \,$ depends on 
$\,x_1,x_2.\,$ Let us also write their iterated version
\begin{equation} (p_{10}^2-E_1^2)\Psi=0, \qquad (p_{20}^2-E_2^2)\Psi=0  \label{8}\end{equation}  with
\begin{equation} E_i=\sqrt{h_i^2}=(\vec p_i^2+m_i^2)^{1\over 2}.   \label{9}\end{equation}  Interesting combinations can be
obtained from the sum and differences of the equations (\ref{7}) or of the iterated equations (\ref{8}):
\begin{equation}(P_0-S)\Psi=0, \qquad (p_0-s)\Psi=0,  \label{10}\end{equation} 
\begin{equation}H_0\Psi=0, \qquad (p_0-\mu)\Psi=0 \label{11}\end{equation}  with
\begin{equation}H_0=2[(p_1^2-m_1^2)+(p_2^2-m_2^2)]_{p_0=\mu}\,\,=\,\,P_0^2-2(E_1^2+E_2^2)+4\mu^2,  \label{12}\end{equation} 
\begin{equation}\mu ={1\over 2P_0}(E_1^2-E_2^2)={1\over 2P_0}(h_1^2-h_2^2)={sS\over P_0}.  \label{13}\end{equation}
 
\subsection{3D reduction of the two-fermion Bethe-Salpeter equation.} The free
propagator $G_0$ will be approached by a carefully chosen expression
$G_\delta$, combining a constraint like $\delta(p_0\! -\!\mu)$ fixing the relative energy, and a global 3D propagator like $-2i\pi
(P_0\! -\! S)^{-1}\beta_1\beta_2$. The argument of the $\delta$ and the inverse of the propagator should be combinations of the
operators used in the free equations (at last approximately and for the positive-energy solutions). Ther exists an infinity of
possible combinations \cite{3,7,14,16,29,30,32,33,34,35,36,37,39,40,41,42,43}. The best choice depends on the quantities one wants
to compute (energy of the lowest state, hyperfine splitting, recoil of a nucleus, etc...) and on the properties one wants to
preserve exactly in the first approximation (cluster separability, Lorentz invariance, heavy mass limits,
 charge conjugation symmetry...). All choices would be equivalent if all correction terms could be computed, but this is of course
impossible. \par We shall see below that the reduced wave function $\Psi$ (from which the relative time-energy degree of freedom can
be trivially eliminated) is given by
\begin{equation}\Psi=G_\delta G_0^{-1}\Phi. \label{14}\end{equation}  The first choice to be made in the 4D $\to$ 3D reduction
is that of the constraint which fixes the 3D hypersurface ($p_0\! =\!\mu$ for example) on which we want to work. The remaining
of
$G_\delta$ is a purely 3D operator, the different choices of which result in different 3D operators applied on a common basic
$\Psi$, and in different rearrangements of a common reduction series giving the 3D potential. The various 3D reductions of the
literature can thus be classified according to the constraint they use. Once this constraint chosen, we can only write different
equivalent forms (or sometimes projections) of the same 3D equation. It is the unavoidable truncation of the reduction series which
makes the difference (numeric, if we simply want to compute predictions of field theory; more fundamental, if we want to write
constraint theory equations with QCD inspired potentials).\par Two natural choices for the constraint are $\delta(p_0\! -\! s)$,
based on the first-order equations (\ref{10}) and  $\delta(p_0\! -\!\mu)$, based on the second-order equations (\ref{11}).
In the pure two-fermion problem, this second choice has the advantage to fix the relative energy in the rest frame to a simple
$\,P_0-$depending number instead of a complicated operator. Other constraints could also be chosen, such as that of Gross
\cite{3} , which puts one particle (normally the heaviest) on mass shell. We made a  nonexhaustive review in ref.\cite{7} . 
\par  For a given $\, P_0$, near $\,P_0\!=\!S,$  the free propagator 
\begin{equation}G_0={1\over {1\over 2}P_0+p_0-h_1+i\epsilon h_1}
\,{1\over {1\over 2}P_0-p_0-h_2+i\epsilon h_2}\beta_1\beta_2  \label{15}\end{equation}  considered as a distribution in $\,p_0
,$  favors the $\,p_0\!=\!s \,$ (or the $\,p_0\!=\!\mu \,$) region. We can thus approach $\,G_0(p_0)\,$ with
$\,G_\delta(p_0)\,$ defined by
\begin{equation}G_\delta(p_0)=\delta(p_0\! -\! s)\,G_S \label{16}\end{equation} or
by\begin{equation}G_\delta(p_0)=\delta(p_0\! -\! \mu)\,G_S
\label{17}\end{equation} with 
\begin{equation}G_S=\int dp_0 G_0(p_0)={-2i \pi\tau \over P_0-S }\,\,   \beta_1 \beta_2  \label{18}\end{equation} 
\begin{equation}\tau={1\over 2} (\tau_1 + \tau_2), \qquad \tau_i = {h_i \over \sqrt{h_i^2}} = {h_i \over E_i} = {\rm sign} (h_i). 
\label{19}\end{equation}  or
\begin{equation}\tau=\Lambda^{++}-\Lambda^{--}, \qquad \Lambda^{ij}=\Lambda_1^i\Lambda_2^j,
\qquad \Lambda_i^\pm={E_i\pm h_i\over 2E_i}. \label{20}\end{equation}  This operator $\tau$ has a clear meaning in the basis
built with the free solutions: it is +1 for $h_1,h_2>0$, -1 for $h_1,h_2<0$ and zero when they have opposite signs. It comes
from the dependence of the $p_0$ integral on the signs of the $i\epsilon h_i$. The denominator of
$G_\delta$ must also contain an infinitesimal imaginary part, obtained by replacing $P_0$ by 
$P_0+i\epsilon P_0$, as usual. The operator $\tau$ insures a common sign for
$p_{01}$, $p_{02}$ and $P_0$, so that there is no sign ambiguity in the sum of the imaginary parts.\par There exists of course an
infinity of possible choices for $G_\delta$. They must however be (at least approximately) identical on the positive energy mass
shell $P_0=E_1+E_2$. An obvious simplification can be made by replacing $\tau$ by 1 (its value for the physical free solutions) or by
$\epsilon(P_0)$ (to get  the correct value for the corresponding antiparticle states too). The merits of the $\tau$ or no-$\tau$
choice are the matter of an old debate.  The operator $\,\tau\,$ brings apparently useless complications in the pure two-body case.
In the two-body plus potential problem, however, the generalization of this operator prevents the "continuum dissolution" disease
(see below). We shall refer to (\ref{18}) as Salpeter's propagator \cite{16}  and to this same expression without the
operator
$\tau$ as the Breit propagator \cite{17} . The Breit propagator is thus:
\begin{equation}G_B=-2i\pi {1\over P_0-S}\,\beta_1\beta_2.\, \label{21}\end{equation}  For a maximum ease of calculation with
the zero-order approximation, one can choose the 3D propagator of Schr\"odinger. Many important physical properties are however
lost at this approximation and can only be recovered by the inclusion of higher-order contributions. By contrast, the 3D
propagator of Sazdjian, based on the second-order equations (\ref{11}), is more complicated but leads to a covariant
$\,G_{\delta}\,$ when combined with the $\,\delta(p_0\!-\!\mu)\,$ constraint \cite{18,7,14}. In the remaining of this
section we shall work with the approached propagator
\begin{equation}G_{\delta}\,=\,\delta(p_0\!-\!s)\,A\,G_B\label{22}\end{equation}
where $\,A\,$  can be $\,\tau\,$ (Salpeter), 1 (Breit) or $\,\Lambda^{++}.$     \par
 We shall write the free propagator as the sum of the approached propagator, plus a rest:
\begin{equation}G_0=  G_{\delta}+G_R.  \label{23}\end{equation}  The Bethe-Salpeter equation  becomes then the inhomogeneous
equation
\begin{equation}\Phi=G_0K\Phi=(G_\delta +G_R)K\Phi=\Psi +G_RK\Phi, \label{24}\end{equation}  with
\begin{equation}\Psi=G_\delta K\Phi \qquad (=G_\delta G_0^{-1}\Phi). \label{25}\end{equation}  Solving (formally) the
inhomogeneous equation (\ref{24}) and putting the result into (\ref{25}), we get
\begin{equation}\Psi=G_\delta K(1-G_RK)^{-1}\Psi=G_\delta K_T\Psi   \label{26}\end{equation}  where
\begin{equation}K_T=K(1-G_RK)^{-1}=K+KG_RK+...=(1-KG_R)^{-1}K  \label{27}\end{equation}  obeys
\begin{equation}K_T=K+KG_RK_T=K+K_TG_RK. \label{28}\end{equation}  The reduction series (\ref{27}) re-introduces in fact the
reducible graphs into the Bethe-Salpeter kernel, but with $G_0$ replaced by $G_R$. Equation (\ref{26}) is a 3D equivalent of the
Bethe-Salpeter equation. It depends on the choice of
$G_\delta.$ \par
 The relative energy dependence of eq. (\ref{26}) can be easily eliminated:
\begin{equation}\Psi=\delta(p_0\! -\! s)\,\psi \label{29}\end{equation} 
and $\,\psi\,$ obeys:
\begin{equation}\psi\,={A\over P_0-S}\,V\,\psi\label{30}\end{equation} where $\,V\,$ is proportional to $\,K_T\,$ with the
initial and final relative energies fixed to $\,s:\,$
\begin{equation}V\,=-2i\pi\,\beta_1\beta_2K_T(s,s).\label{31}\end{equation}
In less compact but more precise notations:    
\begin{equation}\beta_1\beta_2K_T(s,s)\,\equiv\,\int dp_0' dp_0 \delta(p'_0\! -\!s)\beta_1\beta_2K_T(p_0',p_0)\delta(p_0\!
-\!s).\label{32}\end{equation}
Note that we write $\,(p'_0,p_0)\,$ but $\,(s,s),\,$ as we keep $\,s\,$ in operator form.\par   
When $\,A=1\,$ the operator $\,AV\,$ is hermitian. When $\,A=\Lambda^{++}\,$ the operator $\,AV\,$ is hermitian in the
$\,\Lambda^{++}\!=\!1\,$ subspace, and we can write 
\begin{equation}\psi\,={1\over P_0-S}\,\Lambda^{++} \,V\,\Lambda^{++} \psi.\label{33}\end{equation}
When $\,A=\tau\,$ the operator $\,AV\,$ is hermitian in the
$\,\tau^2\!=\!1\,$ subspace, and we can write
\begin{equation}\psi\,={1\over P_0-S}\,\tau \,V\,\tau^2 \psi.\label{34}\end{equation}
Let us recall that the potential $\,V\,$ depends on $\,G_{\delta}\,$ and is thus not the same for the different choices of
$\,A.$\par
These 3D reductions can also be described in terms of transition operators. The 4D transition operator is
\begin{equation}T\,=\,K\,+\,K\,G_0\,K\,+\,\cdots\label{35}\end{equation}
and the 3D transition operator is, when $\,A=\Lambda^{++},\,$ for example
\begin{equation}T^{3D}\,=\,\Lambda^{++}V\Lambda^{++}\,+\,\Lambda^{++}V\Lambda^{++}\,{1\over
P_0-S+i\epsilon}\,\Lambda^{++}V\Lambda^{++}\,+\,\cdots\label{36}\end{equation}
The relation between these two transition operators is simply
\begin{equation}T^{3D}\,=\,-2i\pi\,\beta_1\beta_2\,<T>,\label{37}\end{equation} with
\begin{equation}<T>\,\equiv\,\beta_1\beta_2\,\Lambda^{++}\,\beta_1\beta_2\,T(s,s)\,\Lambda^{++}.\label{38}\end{equation}
We have indeed
$${-1\over2i\pi}\,\beta_1\beta_2\,T^{3D}\,=\,\,<K_T>\,+\,<K_T>\,{-2i\pi\over
P_0-S+i\epsilon}\,\beta_1\beta_2\,<K_T>\,+\cdots$$  
$$=\,\,<K_T>\,+\,<K_T\,G_{\delta}\,K_T>\,+\cdots\,=\,\,<K_T\,(1-\,G_{\delta}\,K_T\,)^{-1}>$$
$$=\,\,<K\,(1-G_R\,K)^{-1}\,(1-\,G_{\delta}\,K\,(1-G_R\,K)^{-1}\,)^{-1}>$$
\begin{equation}=\,\,<K\,(1-G_R\,K-\,G_{\delta}\,K\,)^{-1}>\,\,=\,\,<K\,(1-G_0\,K\,)^{-1}>\,\,=\,\,<T>.\label{39}\end{equation}
A Bethe-Salpeter equation leading directly to the same 3D reduction can be obtained by replacing  the kernel $\,K\,$ by the
instantaneous kernel $\,<K_T>.\,$ The corresponding transition operator becomes in this case
\begin{equation}T\,=\,\,<K_T>\,+\,<K_T>\,G_0\,<K_T>\,+\cdots\,=\,\,<T>.\label{40}\end{equation}   
\par In contrast with the textbook cases, the potentials deduced from field theory are in general energy-dependent (although
one starts often with an energy-independent approximation). The eigenvalue of the hamiltonian is then a function of the energy itself, and
the energy spectrum is given by the solutions of the algebric equations
\begin{equation}E\,=\,E_{\lambda}\,(E)\label{41}\end{equation} but the corresponding wave functions are no more orthogonal with the
the usual scalar product, so that the usual methods of perturbation calculation have to be revisited \cite{19}\par
 Until now we did not specify a reference frame, and our equations are not manifestly covariant
under the Lorentz group. The consequences of this must however be carefully discussed.\par
-- Trough not manifestly covariant, equations like (\ref{26}) can always be transformed back into the original Bethe-Salpeter
equation, provided the reduction series (\ref{27}) is not truncated. The inclusion of higher-order terms will thus, in
principle, improve an approached covariance. Beyond this somewhat trivial consideration, it would of course be interesting to be
able to simplify the equations in a way which preserves the covariance. If we use the covariant two-body Sazdjian propagator 
 \cite{18,7,14}, based on the second-order equations (\ref{11}), instead of Breit's, it becomes possible to truncate the
reduction series in a covariant way (one might for example keep only the first term of the reduction series at the ladder
approximation, i.e. the Born term).    
\par -- Even truncated, our equations could be made formally invariant by introducing a unit vector
$\,n=(1,\vec 0)\,$ (we can call this vector the laboratory time unit) and by making all elements in our equations invariant by
using scalar products with
$\,n.$ If we want more than a formal invariance, this unit vector can however not be external to the system: one must therefore
define
$\,n\,$ as
$\,P/\sqrt{P^2}\,$ (the time unit of the center of mass reference frame), assuming $\,P^2>0.$ \par
-- It must be noted that our non-covariant equations
admit, among others, $\,\vec P=0\,$ solutions which are identical to the $\,P^2>0\,$ solutions of the covariant equations in the
center of mass reference frame. We could speak of "weak covariance".\par 
-- An equation like (\ref{33}) can be assumed to
have been written in the center of mass reference frame and made explicitly covariant by using the vector $\,P/\sqrt{P^2}.$ This
covariant form could be used to write the system of equations in other reference frames. There is no reason to do that in the pure
two-body case, but it will become necessary when other objects (as an external potential or a third particle) are present. The
 $\, G_B\,\delta(p_0-s)\, $ combination written in the center of mass reference frame looses its simplicity when
covariantized.\par
The cluster separability property is clearly satisfied, as the
"switching off" of the mutual interaction leads to a pair of free Dirac equations.
 
\section{The three-fermion problem.}

\subsection{A cluster separable 3D equation.}
A three-fermion 3D equation, inspired by the two-fermion equation (\ref{30}) with $\,A=1\,$ could be
\begin{equation}\psi\,={1\over P_0-S}\,(\,V_{12}\,+\,V_{23}\,+\,V_{31}\,)\,\psi\label{42}\end{equation} 
with $\,S\!=\!h_1+h_2+h_3.\,$ The potentials are the two-body potentials defined by (\ref{31}). Each two-body potential
$\,V_{ij}\,$ was a function of the partial energy $\,P_{ij0},\,$ which we shall fix to its (ij)+k cluster-separated limit
$\,P_0\!-\!h_k.\,$ At the
$\,V_{23}=V_{31}=0\,$ limit, for example, we shall thus get two completely independent equations:
\begin{equation}\bigg[\,P_{120}-h_1-h_2\,\bigg]\,\psi_{12}\,=\,
V_{12}(P_{120})\,\psi_{12},\qquad p_{30}\,\psi_3\,=\,h_3\,\psi_3\label{43}\end{equation}
with 
\begin{equation}\psi=\psi_{12}\,\psi_3,\qquad P_0\,=\,P_{120}\,+\,p_{30}.\label{44}\end{equation}
Our 3D equation (\ref{42}) satisfies thus clearly the cluster separability requirement. Furthermore, the three cluster-separated
limits are exact equivalents of the corresponding two-fermion Bethe-Salpeter equations.\par In the two-body problem all
quantities can be defined in the center of mass reference frame. In the two-body plus potential problem we had to start in the
laboratory frame but to consider also the center of mass reference frame at the no-external potential limit. In the three-body
problem we must start in the three-body center of mass frame (unless we are satisfied with a "weak covariance") and consider the
center of mass reference frames of the three possible two-body subsystems obtained by cluster separation. 
The fact that the three cluster-separated limits are exact insures in principle the Lorentz invariance / cluster separability
requirement: at the cluster separated limits the two-body equation can indeed be transformed back into a covariant two-body
Bethe-Salpeter equation. There is no necessity of introducing Lorentz boosts by hand.\par
 Our two-body potentials are the sum of an infinity of contributions symbolized by Feynman graphs. Keeping only  the first
one (Born approximation) or a finite number of them renders the Lorentz covariance of the two-fermion clusters only approximate. A Born
approximation preserving the  Lorentz invariance / cluster separability property can be obtained by using  another 3D reduction based on a
covariant second-order two-body propagator of Sazdjian \cite{18,7,14} (combined with a covariant substitute of $\,\Lambda^{++}\,$
to prevent continuum dissolution - see below). This leads to a 3D three-cluster equation which is  covariantly Born approximable,
but more complicated \cite{20}.\par
    
\subsection{The continuum dissolution problem.} Unfortunately, our 3D equation (\ref{42}) suffers of continuum dissolution.  It
is  indeed possible to build a continuum of solutions with any a priori given total energy by combining asymptotically free
fermions with opposite energy signs. Any physical bound state is thus  degenerate with such a continuum and the building of
normalisable bound state wave functions becomes impossible \cite{8,9,10,11}. In the pure two-body case the  energy of a system
$\,(+,-)\,$ in the total rest frame is 
\begin{equation}E_1-E_2=\,\sqrt{\vec p^2+m_1^2}-\sqrt{\vec p^2+m_2^2}\,=\,{m_1^2-m_2^2\over \sqrt{\vec p^2+m_1^2}+\sqrt{\vec
p^2+m_2^2}}\label{45}\end{equation}
and lies thus between $\,m_1-m_2\,$ (whichever the sign) and zero.  We have thus no problem if  we make the
 assumption that the energies of the bound states lie between
$\,\vert\, m_1-m_2\vert\,$ and $\,m_1+m_2.$ Below $\,\vert\, m_1-m_2\vert\,$ we meet the "strong field" problem. The strong
field problem for the one-body plus potential and two-body sustems, the continuum dissolution problem for the
two-body plus potential and three-body systems are both consequences of the possibility of pair creation.\par 
In the three-body case, the mixing
with asymptotically separated (12)(3) subsystems of opposite energy signs is excluded by the total momentum conservation, as in the
two-fermion case. For the mixing with three-fermion asymptotically free states, let us consider for example
\begin{equation}E_3-E_1-E_2=\,\sqrt{\vec p_3^2+m_3^2}-\sqrt{\vec p_1^2+m_1^2}-\sqrt{\vec
p_2^2+m_2^2\,},\qquad\vec p_1+\vec p_2+\vec p_3=\vec 0.\label{46}\end{equation}
 There is no lowest value (we can for example have $\,\vec p_3\!=\!0\,$ and $\,\vec p_1\!=\!-\vec p_2\,$ arbitrarily large). 
The highest values are obtained when $\,\vec p_1\,$ and $\,\vec p_2\,$ have the same direction, and for $\,\vec
p_1/m_1\!=\!\vec p_2/m_2\!=\!-\,\vec p_3/(m_1\!+\!m_2).\,$ In this case we have
\begin{equation}E_3-E_1-E_2=\,\sqrt{\vec p_3^2+m_3^2}-\sqrt{\vec p_3^2+(m_1+m_2)^2}\label{47}\end{equation}
so that $\,E_3-E_1-E_2\,$ lies finally between $\,-\infty\,$ and $\,(m_3\!-\!m_1\!-\!m_2)\theta(m_3\!-\!m_1\!-\!m_2).\,$
Symmetrically, $\,E_1+E_2-E_3\,$ lies between  $\,(m_1\!+\!m_2\!-\!m_3)\theta(m_3\!-\!m_1\!-\!m_2)\,$ and $\,\infty.\,$ If we
assume that the energies of the bound states lie between $\,(m_1\!+\!m_2\!+\!m_3\!-2\, \hbox{Inf}(m_i))\,$ (i.e. above the highest
negative-energy threshold) and $\,(m_1\!+\!m_2\!+\!m_3),\,$ there is no degenerescence with the "one plus-two minus" states.
On the contrary, no weak field assumption could prevent the degenerescence with the "two plus-one minus" states.\par
       This continuum dissolution problem can be cured by simply introducing the products of noncovariant projectors
$\,\Lambda^{++}_{ij}=\Lambda^+_i\Lambda^+_j\,$ (on the eigenstates of the free $\,h_i,h_j\,$ with positive eigenvalues) into the
corresponding $\,G_{\delta ij},\,$ defining the two-body potentials, and around these potentials.  This changes finally equation
(\ref{42}) into 
\begin{equation}\psi={1\over P_0-S}\,(\,
\Lambda^{++}_{12}V_{12}\Lambda^{++}_{12}\,+\,\Lambda^{++}_{23}V_{23}\Lambda^{++}_{23}\,+\,
\Lambda^{++}_{31}V_{31}\Lambda^{++}_{31}\,)
\,\psi\label{48}\end{equation}
built with the three two-fermion equations (\ref{33}). 
There exist solutions of eq.(\ref{48}) which are entierely contained in the (+++) subspace. These solutions obey
an equation with a spinless Salpeter free term:
\begin{equation}(P_0-E_1-E_2-E_3)\,\psi=\,\Lambda^{+++}\,(\,
V_{12}\,+\,V_{23}\,+\,
V_{31}\,)\,\Lambda^{+++}
\,\psi.\label{49}\end{equation}
It is also possible to build a three-fermion equation with three Salpeter equations:
\begin{equation}\psi={1\over P_0-S}\,(\,
\tau_{12}V_{12}\tau^2_{12}\,+\,\tau_{23}V_{23}\tau^2_{23}\,+\,\tau_{31}V_{31}\tau^2_{31}\,)
\,\psi.\label{50}\end{equation}
This equation admits solutions lying entierely in the $\,(+++)\oplus(+--)\oplus(-+-)\oplus(--+)\,$ subspace.
 The
remarks above about the cluster separability and Lorentz invariance of equation (\ref{42}) remain true for equations
(\ref{48}) and (\ref{50}).\par
In the remaining of this work we shall adopt the positive-energy equation (\ref{48}) as our basic three-cluster 3D equation.

\section{Heavy mass limits and external potentials.}

\subsection{Two fermions in an external potential.}

 The two-fermion plus potential problem can be approached in two ways which we shall call the two-body and the three-body
approaches. In the two-body approach, the external potential is included in the definition of the creation and
annihilation operators of field theory. Practically, we can keep the equations obtained in the treatment of the pure
two-fermion problem,  using simply a generalized definition of the
$\,h_i$  \cite{9,10,11}:
\begin{equation}h_i = \vec \alpha_i\, . \vec p_i + \beta_i\, m_i\,+\,V_i(\vec x_i)\label{51}\end{equation} where $\,V_i\,$ is
the external potential acting on the fermion i. 
 All quantities can be expanded on the basis built with the eigenstates of
$\,h_1\,$ and $\,h_2.$ In the three-body approach, the free creation and annihilation operators are used and the external
potential is treated as an heavy third body. This approach will be presented as an heavy mass limit of the three fermion problem
in subsection 4.3. Here we shall adopt the two-body approach.\par 

 The equations for two fermions in an external potential must be written in the
laboratory reference frame (in which the external potential is defined) and are not Lorentz invariant. If we
"switch off" the mutual interaction we get a pair of uncoupled equations for two independent fermions in an external potential. If
we switch off the external potential the equations remain written in the laboratory frame, which still refers to the vanished
external potential. If the reduction series is not truncated, this no-external interaction limit of the equation is equivalent to
a pure two-fermion covariant Bethe-Salpeter equation. If the reduction series is truncated, the pure two-body cluster equation is
only "weakly covariant".  If we want a truly cluster separable system, we must then use
equations which are covariant at the vanishing external potential limit, such as Sazdjian's equations with the generalization
(\ref{51}) of the
$\,h_i\,$ \cite{18,7,14}. 
 In the two-body plus potential problem the total spatial momentum is no more conserved and the mixing with the continuum forbids
the building of normalizable solutions. A easy way of seeing how this happens is to try to build these states by perturbations of
the no-mutual interaction equations: the operators $\,h_1, h_2\,$ are then diagonal and the higher-order contributions contain
denominators in $\,P_0-h_1-h_2,$ which can vanish on a continuum. In  Salpeter's equation the operator
$\,\tau\,$ kills the matrix elements of $\,\tau\, V\,$ between the $\,\tau^2=1\,$ states $\,(+,+),(-,-)\,$ and the  $\,\tau^2=0\,$
states $\,(+,-),(-,+)\,$ in the basis built with the eigenstates of $\,(h_1,h_2),$ forbidding the mixing.
\subsection{Heavy mass limits in the two-fermion problem.}
When the mass of one of the fermions goes to
infinity, we must find the equation of the other fermion in a potential (a Dirac equation with a Coulomb potential in QED).
This limit can be obtained directly, or via a rearrangement of the equation. With Breit's equation with the second-order
constraint and with Sazdjian's equation, the higher-order ladder and crossed terms cancel mutually at the one-body limit, so
that the correct limit of the potential is already contained in the Born term. With Salpeter's equation with a second-order
constraint, the limit is also a Dirac-Coulomb equation, but solved with respect to the
$\,\Lambda^+\psi\,$ part of the wave function: the
$\,\Lambda^-\psi\,$ part is transformed into higher-order contributions to the potential. The physical content remains identical
to that of a Dirac-Coulomb equation, but the correct limit of the potential is no more entierely contained in the Born term, so
that a truncation of the potential would spoil the one-body limit \cite{3,4,5,6,7}.

\subsection{Heavy mass limits in the three-fermion problem.}
 When $\,m_3,$ for example, goes to infinity (two-body limit), we get, writing $\,P_0=W_{12}+m_3\,$ in our basic three-cluster
equation:
\begin{equation}W_{12}\,\psi\,=\,(h_1+h_2+\Lambda^{++}_{12}\,V_{12}\,\Lambda^{++}_{12}+\Lambda^+_2\,V^+_2\,\Lambda^+_2
+\Lambda^+_1\,V^+_1\,\Lambda^+_1)\,\psi\label{52}\end{equation}
with $\,(P_0-h_3)\,$ replaced by $\,W_{12}\,$ in $\,V_{12}.\,$ The
potential
$\,V^+_2\,$ is given by the series
\begin{equation}V^+_2\,=\,V_2\,+\,V_2\,{\Lambda^-_2\over W_{12}-h_1-h_2}\,V_2\,+\,...\label{53}\end{equation}
where $\,V_2\,$ is an external potential acting on fermion 2 (it is a Coulomb potential in QED, if we use the second-order
constraint (\ref{17}),  but in general it could still depend on $\,(W_{12}-h_1)\,$). The potential
$\,V^+_1\,$ is given by a similar formula.\par\noindent The differences with the two-body approach of the two-body plus potential
problem are:\par\hfill\break
\indent -- The projectors $\,\Lambda^{\pm}_i\,$ are the free ones.\par
-- We have now projectors around the external potential terms $\,V^+_2,V^+_1.$\par
-- These external potential themselves are now given by the series (\ref{53}).\par
-- $\,V_2\,$ could still depend on $\,(W_{12}-h_1)\,$ and vice-versa. \par\hfill\break\noindent
The cluster separability property in the three-body way survives the high-mass limit: switching off $\,V_2\,$ and $\,V_1\,$ leads
to the equation for the (12) two-fermion system, switching off $\,V_{12}\,$ and $\,V_1\,$ leads to a free Dirac equation for
fermion 1 with the $\,\Lambda^+_2\,$ projection of the Dirac equation for fermion 2 in the external potential
$\,V_2.\,$ In the two-body approach of the two-fermion plus potential problem we had however more: switching off only the mutual
interaction led to a pair of independent equations for each fermion in the external potential. Here, in the
three-body approach, the equation does not split perfectly into two parts: we can write $\,W_{12}\!=\!W_1+W_2,\,$ but we have 
$\,W_{12}-h_1\,$ instead of $\,W_2\,$ in the series defining $\,V^+_2\,$ and vice-versa. This is a consequence of the energy
dependence introduced into (\ref{53}) by the use of the anti-continuum dissolution projectors, combined with our choice of
putting the spectator fermion on the mass shell in each two-body interaction, and neglecting the three-body terms which could
balance this modification. The discrepancy is of order
$\,V^4.\,$ At the two-fermion plus potential level, a better but still not perfect equation would be obtained by replacing $\,h_1\,$
by
$\,h_1+V_1\,$ in the series defining $\,V^+_2\,$ and vice-versa.

\section{Towards a perturbation calculation.}
\subsection{Bethe-Salpeter equivalent of our basic three-cluster equation. }
The three-fermion Bethe-Salpeter equation can be written
\begin{equation}\Phi=\left[G_{01}G_{02}K_{12}+G_{02}G_{03}K_{23}+G_{03}G_{01}K_{31}+
G_{01}G_{02}G_{03}K_{123}\right]\Phi\label{54}\end{equation}
where
$K_{123}$ is given by the sum of the purely three-body irreducible contributions. We would like to get our basic three-cluster 3D equation
by approximating the three-fermion Bethe-Salpeter equation, and to possibly recover the neglected contributions in a perturbation
calculation afterwards. A three-fermion Bethe-Salpeter equation with instantaneous (i.e. independent of the relative energies)
and positive-energy kernels can be transformed into a 3D Salpeter equation \cite{10}.   We shall thus neglect the three-body
kernel and replace the two-body kernels by instantaneous positive-energy kernels (with the spectator fermions on their mass shell)
which are equivalent at the cluster-separated limits:
\begin{equation}K_{12}\,\approx\,\,
<K_{T12}(P_0-h_3)>\,\,=\,
\,{-1\over2i\pi}\,\beta_1\beta_2\,\Lambda^{++}_{12}\,V_{12}(P_0\!-\!h_3)\,\Lambda^{++}_{12}\,,\cdots 
.\label{55}\end{equation}     The Bethe-Salpeter equation becomes
$$\Phi={-1\over2i\pi}\,G_{01}G_{02}G_{03}\,\beta_1\beta_1\beta_3\left[\,\Lambda^{++}_{12}\,V_{12}\,\Lambda^{++}_{12}\,\psi_{12}\,+\,
\Lambda^{++}_{23}\,V_{23}\,\Lambda^{++}_{23}\,\psi_{23}\right.$$
\begin{equation}\left.\,+\,\Lambda^{++}_{31}\,V_{31}\,\Lambda^{++}_{31}\,\psi_{31}\,\right]\label{56}\end{equation}
where 
\begin{equation}\psi_{ij}(p_{k0})=\beta_k\,G_{0k}^{-1}\int dp_{ij0}\,\Phi.\label{57}\end{equation}
This leads to a set of three coupled integral equations:
$$\psi_{12}(p_{30})={-1\over2i\pi}\int
dp_{120}\,G_{01}G_{02}\,\beta_1\beta_2\left[\,\Lambda^{++}_{12}\,V_{12}\,\Lambda^{++}_{12}\,\psi_{12}(p_{30})\,\right.$$
\begin{equation}\left.+\,
\Lambda^{++}_{23}\,V_{23}\,\Lambda^{++}_{23}\,\psi_{23}(p_{10})\,+\,\Lambda^{++}_{31}\,V_{31}\,
\Lambda^{++}_{31}\,\psi_{31}(p_{20})\,\right] \label{58}\end{equation}
where $\,p_{10},p_{20}\,$ must be written in terms of $\,P_0,p_{30},p_{120}:$   
\begin{equation}p_{10}\,=\,{P_0-p_{30}\over2}\,+\,p_{120},\qquad p_{20}={P_0-p_{30}\over2}\,-\,p_{120}.
\label{59}\end{equation}
and similarly for $\,\psi_{23}\,$ and $\,\psi_{31}.\,$  We shall now search for solutions $\,\psi_{ij}(p_{k0})\,$   
 analytical in the lower Im$(p_{k0})\!<\!0\,$ half plane and  perform the integration (\ref{58})
 by closing the integration paths around these regions.  
The only singularities will then be the poles of the free propagators. The result is
$$\psi_{12}(p_{30})=\,{\Lambda^{++}_{12}\over
(P_0-S)-(p_{30}-h_3)+i\epsilon}\,\left[\,\Lambda^{++}_{12}\,V_{12}\,\Lambda^{++}_{12}\,\psi_{12}(p_{30})\right.$$
\begin{equation}\left. +\,
\Lambda^{++}_{23}\,V_{23}\,\Lambda^{++}_{23}\,\psi_{23}(h_1)\,+\,\Lambda^{++}_{31}\,V_{31}\,\Lambda^{++}_{31}\,
\psi_{31}(h_2)\,\right]\,,\cdots .\label{60}\end{equation}
Using these equations to compute the projections $\,-2i\pi\,\Lambda^+_k\, \psi_{ij}(h_k)\,$ we find that these three
expressions are equal (let us call them $\,\psi$) and obey the $\,\Lambda^{+++}\,$ projection of our three-cluster equation
(\ref{48}):    
\begin{equation}\psi\,=\,{\Lambda^{+++}\over P_0-S}\, \left[\,V_{12}\,+\,
V_{23}\,+\,V_{31}\,\right]\,\Lambda^{+++}\,\psi.\label{61}\end{equation}
Furthermore, we have also
$$\int\! dp_0\Phi\equiv\int\!dp_{10}dp_{20}dp_{30}\,\delta(p_{10}+p_{20}+p_{30}-P_0)\,\Phi=\,\int\!
dp_{k0}\,G_{0k}\,\beta_k\,\psi_{ij}(p_{k0})\,$$
\begin{equation}=\,-2i\pi\,\Lambda^+_k\, \psi_{ij}(h_k)\,=\psi.\label{62}\end{equation}
Solving (\ref{60}) with respect to $\,\psi_{12}(p_{30})\,$ gives  
\begin{equation}\psi_{12}(p_{30})=\,{-1\over2i\pi}\,{\Lambda^{+++} \over
(P_0-p_{30})-(S_{12}+\Lambda^{++}_{12} V_{12} \Lambda^{++}_{12}
)+i\epsilon}\,\left[\,V_{23}\,+\,V_{31}\,\right]\,\Lambda^{+++}\,\psi\label{63}\end{equation}
which confirms the above assumption about the analyticity of the $\,\psi_{ij}(p_{k0}).\,$ 

\subsection{Relation with Faddeev formalism and Gross' spectator model.}
The nonrelativistic Faddeev equations can be obtained by transforming Schr\"odinger's equation (in a first step, without
three-body terms) into a set of three coupled equations for three parts $\,T_{ij}$ of the transition operator
($\,T_{ij}$ denotes the contribution of all graphs beginning by a (ij) interaction). The input is the set of the three
two-body transition operators and the resulting series expansion contains only connected graphs (never twice the same
two-body transition operator). This formalim is well adapted to the description of the various
scattering processes, such as
$\,(12)+3\to (12)+3
$ (elastic scattering),
$\,(12)+3\to (12)^*+3 $
 (excitation), $\,(12)+3\to 1+(23) $ (rearrangement), $\,(12)+3\to 1+2+3 $ (breakup). The (123) bound states correspond to the
poles of this transition operator. \par
The structure of these equations can be generalized to relativistic equations which are not necessarily (exactly) reducible to
a single 3D equation. \par
In Gross'  spectator model \cite{12,13}, Faddeev's type equations are deduced from the Bethe-Salpeter equation and
the transition operator is written in terms of the two-body transition operators. The relative time variables are then eliminated
by putting, in each three-body  propagator, all the "offmassshallness" on the only fermion which interacts before and after.  The
Lorentz invariance-cluster separability requirement is satisfied by applying suitable Lorentz boosts on the two-body transition
operators.
\par The main difference between our approach and Gross' approach comes from the fact that we are (presently) interested
by the (123) bound states only, to be computed principally by using a single 3D equation (such as the one from which Faddeev starts).
Instead of working with the two-body transition operators, we work with the two-body potentials. We can however present our
approach in terms of the two-body transition operators. Our approximation of each two-body transition operator is then unique and
does not depend on the operators which come in front and behind it in the expansion of the Faddeev equations. This enables us to
describe our model by a single 3D potential equation, by a set of three Faddeev equations, or by the expansion of the three-body
transition operator in terms of the two-body ones. Another option of our model is to not introduce Lorentz boosts by hand: the
Lorentz-invariance / cluster separability requirement is exactly satisfied if we do not truncate our potentials (practically, this
means that the satisfaction of this requirement and the approximation of the potentials can be improved together).  Let us go back
to the three-fermion Bethe-Salpeter equation  and neglect for simplicity the three-body irreducible kernel $\,K_{123}. \,$ Defining
\begin{equation}\Phi_{12}\,=\,G_{01}G_{02}\,K_{12}\,\Phi,\cdots\qquad\,\Phi\,=\,\Phi_{12}\,+\,\Phi_{23}\,+\,\Phi_{31}\,, 
\label{64}\end{equation} we get
\begin{equation}(1\,-\,G_{01}G_{02}\,K_{12})\,\Phi\,=\,\Phi_{23}\,+\,\Phi_{31}  \label{65}\end{equation}
\begin{equation}\Phi\,=\,(1\,+\,G_{01}G_{02}\,T_{12})\,(\,\Phi_{23}\,+\,\Phi_{31}\,)  \label{66}\end{equation}
\begin{equation}\Phi_{12}\,=\,G_{01}G_{02}\,T_{12}\,(\,\Phi_{23}\,+\,\Phi_{31}\,).  \label{67}\end{equation}
Writing then
\begin{equation}\Phi_{12}\,=\,G_{01}G_{02}G_{03}\,\beta_1\beta_2\beta_3\,\chi_{12},\,\cdots\label{68}\end{equation}
in order to factor out the propagators, we get
\begin{equation}\chi_{12}\,=\,\beta_1\beta_2\,T_{12}\,G_{01}G_{02}\,\beta_1\beta_2\,(\,\chi_{23}\,+\,\chi_{31}\,).
\label{69}\end{equation}\hfill\break  
The (12) transition matrix element corresponding to our approximation (\ref{55}) of $\,K_{12}\,$ is
$$T_{12}(\,p'_{120},\,p_{120},\,P_0-p_{30}\,)\,\approx\,T_{12}^0(p_{30})\,$$
\begin{equation}\,\,\equiv\,\,\,<K_{T12}(P_0-h_3)>
(1-G_{01}G_{02}<K_{T12}(P_0-h_3)>)^{-1}.\label{70}\end{equation}
This $\,T_{12}^0(p_{30})\,$ is  analytical in the
Im$(p_{30})<0\,$ half plane and 
\begin{equation}T_{12}^0(h_3)\,=\,<T_{12}(P_0-h_3)>.\,\label{71}\end{equation}   
This approximation, combined with equation (\ref{69}), implies that $\,\chi_{12}(p_{120},p_{30})\,$ is independent of
$\,p_{120}\,$ (let us write thus $\,\chi_{12}(p_{30})\,$). Equation (\ref{69}) becomes then
\begin{equation}\chi_{12}(p_{30})\,=\,\beta_1\beta_2\,T_{12}^0(p_{30})\,\int d\,p_{120}\,
G_{01}G_{02}\,\beta_1\beta_2\,[\,\,\chi_{23}(p_{10})\,+\,\chi_{31}(p_{20})\,]\label{72}\end{equation}                 
where $\,p_{10},p_{20}\,$ must be written in terms of $\,P_0,p_{30},p_{120}.\,$ Equation (\ref{72}), together with
similar equations for $\,\chi_{23}(p_{10})\,$ and $\,\chi_{31}(p_{20}),\,$ admits solutions which are analytical in the
Im$(p_{k0})<0\,$ half planes. At $\,p_{k0}=h_k,\,$ we get the 3D Faddeev equations
\begin{equation}\chi_{12}(h_3)\,=\,T^{\,3D}_{12}(P_0-h_3)\,{1\over
P_0-S}\,[\,\,\chi_{23}(h_1)\,+\,\chi_{31}(h_2)\,]\,,\cdots \label{73}\end{equation}
which can easily be transformed back into our basic three-cluster potential equation (by performing in the reverse
order the transformations made above at the 4D level). Note that we would get the same result by approaching directly              
$\,T_{12}\,$ by $\,T_{12}^0(h_3)\,$ instead of $\,T_{12}^0(p_{30}).\,$ This means approaching $\, K_{12}\,$ by a given function
$\,K_{12}^0(p_{30}),\,$ analytical in the
Im$(p_{30})<0\,$ half plane and equal to $\,<K_{T12}\,(P_0-h_3)>\,$ at $\,p_{30}=h_3.$ \par
The key point of the manipulations above is the dominance of the positive-energy poles of $\,G_{01}G_{02}\,$ in (\ref{69}).
This was obtained by approaching $\,K_{T12}\,$ (or $\,T_{12}\,$) by a constant with $\,\Lambda^{++}_{12}\,$  positive-energy
projectors. We could try to insure the dominance of these two poles more economically. Let us come back to equation (\ref{69})
without making any approximation on $\,T_{12}.\,$ The elements of (\ref{69}) are then the operator and functions
\begin{equation}T_{12}\,(\,p'_{120},\,p_{120},\, P_0-p_{30})\label{74}\end{equation} 
\begin{equation}\chi_{12}\,(\,p'_{120},\,p_{30}\, ),\quad \chi_{23}\,(\,p_{230},\,p_{10}\,
),\quad\chi_{31}\,(\,p_{310},\,p_{20}\, )\label{75}\end{equation}
and (\ref{69}) must be integrated with respect to $\,p_{120},\,$ with $\,p_{30}\,$ fixed. We must thus write
$\,p_{230},\,p_{10},\,p_{310},\,p_{20}\,$ in terms of $\,p_{120},\,p_{30}.\,$ Searching as above for solutions
$\,\chi_{ij}\,(\,p'_{ij0},\,p_{k0}\, )\,$ which are analytical in the Im$(p_{k0})<0\,$ half planes, we shall
close our integration path clockwise (counterclockwise) in front of $\,\chi_{23}\,$ ($\,\chi_{31}\,$) and keep thus the
pole of $\,G_{01}\,$ ($\,G_{02}\,$), which puts fermion 1 (2) on its positive-energy mass shell. The elements of (\ref{69})
are then replaced by
\begin{equation}T_{12}\,(\,p'_{120},\,p_{120},\, P_0-p_{30})\,\to\,\,
T_{12}\,(\,p'_{120},\,s_{12}\,\mp\,{P_0-S\over2}\,\pm\,{p_{30}-h_3\over2},\,
P_0-p_{30})\,\label{76}\end{equation}                     
\begin{equation}\int d\,p_{120}\,G_{01}G_{02}\,\beta_1\beta_2\,\to\,{-2i\pi\over
(P_0-S)-(p_{30}-h_3)+i\epsilon}\label{77}\end{equation}
\begin{equation}\chi_{23}\,(\,p_{230},\,p_{10}\,
)\,\to\,\chi_{23}\,(\,s_{23}\,+\,{P_0-S\over2}\,-\,(\,p_{30}-h_3\,),\,h_1\,)\label{78}\end{equation}  
\begin{equation}\chi_{31}\,(\,p_{310},\,p_{20}\,
)\,\to\,\chi_{31}\,(\,s_{31}\,-\,{P_0-S\over2}\,+\,(\,p_{30}-h_3\,),\,h_2\,)\label{79}\end{equation}      
where we take the upper signs  in (\ref{76}) in front of $\,\chi_{23},\,$ the lower signs in front of
$\,\chi_{31}.$\par The manipulations above are submitted to some restrictions on the $\,T_{ij}.\,$  In the $\,p_{k0}\,$
variable the $\,T_{ij}\,$ must be analytical in the Im$(p_{k0})<0\,$ half planes. In the $\,p'_{ij0}\,$ and
$\,p_{ij0}\,$ variables the $\,T_{ij}\,$ must be analytical in the whole complex plane. Moreover, the 
$\,T_{ij}\,$ must also be asymptotically bounded in the three variables  and contain $\,\Lambda^{++}_{ij}\,$
positive-energy projectors. These conditions are of course not satisfied by the exact transition matrix elements, but we shall
assume that the singularities and the negative-energy parts of the
$\,T_{ij}\,$ can be neglected in the computation of the integrals with respect to the relative times. \par           If we take
then the equation for
$\,\chi_{12}\,$ at
$\,p_{30}\!=\!h_3\,$ and
$\,p'_{120}=s^{\pm}_{12}=s_{12}\,\pm\,{1\over2}(P_0-S),\,$ etc..., we get a closed  system of six 3D equations:
\begin{equation}\chi^+_{12}\,=\,-2i\pi\,\beta_1\beta_2\,(\,T^{+-}_{12}\,{1\over
P_0-S}\,\,\chi^+_{23}\,+\,T^{++}_{12}\,{1\over P_0-S}\,\,\chi^-_{31}\,)\label{80}\end{equation}
\begin{equation}\chi^-_{12}\,=\,-2i\pi\,\beta_1\beta_2\,(\,T^{--}_{12}\,{1\over
P_0-S}\,\,\chi^+_{23}\,+\,T^{-+}_{12}\,{1\over P_0-S}\,\,\chi^-_{31}\,)\label{81}\end{equation}
and similarly for $\,\chi^{\pm}_{23}\,$ and $\,\chi^{\pm}_{31}.\,$\par
In each $\,T_{ij}\,$ two fermions are on the mass shell: the spectator fermion k and the fermion which is not going to
interact next left ($\,p'_{ij0}\,$) or next right ($\,p_{ij0}\,$). This is of course the philosophy of Gross' spectator
model \cite{12,13}. \par
If we replace $\,s^{\pm}\,$ by $\,s\,$ in the $\,T_{ij}\,$ we get our basic three-cluster model. This supplementary
approximation can be introduced by noticing that we already neglected the contributions of the singularities of
$\,T_{ij}\,$ in the $\,p'_{ij0}\,$ and $\,p_{ij0}\,$ complex planes. The dependence on these variables can thus not be
very strong, as we know that a function which is analytical and bounded on the whole complex sphere must necessarily be
a constant. Using the same argument of the consistency of the approximations, we could also argue that $\,T_{ij}\,$
constant implies $\,K_{Tij}\,$ and $\,K_{ij}\,$ constant with $\,K_{Tij}=K_{ij}.\,$ Our basic three-cluster equation
is thus not a priori a worse approximation of the three-fermion Bethe-Salpeter equation than Gross' equations, nor a better
approximation than the simpler positive-energy instantaneous approximation or the corresponding Born approximation. It reflects
the choice of preserving exactly the exact cluster-separated limits. \par
It is thus important to evaluate the importance of the neglected contributions in a perturbation calculation.         

 \subsection{Correction terms.}
The contributions neglected above in the transformation of the Bethe-Salpeter equation into a three-cluster 3D equation are given at the
Bethe-Salpeter level. We would like to transform them corrections to the 3D equation. However, the exact
Bethe-Salpeter equation has no more the simple analyticity structure in the relative energies which allows the 3D reduction at the
positive-energies instantaneous approximation (\ref{55}).\par
We made several attempts to build a 3D perturbation calculation starting with our basic three-cluster equation. Some of them led
to interesting results, but not to the kind of 3D perturbation terms we wanted. In the two-fermion problem, a perturbation
expansion can be built around an approximation of the propagator (as in section 2), of the kernel (as in \cite{21} for example), or
of both. Our approximation (\ref{55}) for the three-fermion Bethe-Salpeter equation is an approximation of the kernels: the
three-fermion kernel is neglected and the two-fermion kernels are replaced by series based on approximations of the two-fermion
propagators. In order to transform the correction terms, we shall write, instead of (\ref{55})
$$\delta(p'_{30}\!-\!p_{30})\,K_{12}(p'_{120},p_{120},P_0\!-\!p_{30})$$
\begin{equation}\,\approx\,
\delta(p'_{30}\!-\!h_3)\,\Lambda^+_3<K_{T12}(P_0\!-\!h_3)>\,\equiv\,K^0_{12}(p'_{30}).\label{82}\end{equation}
This gives directly the (+++) projection of our basic three-cluster equation for the integral (\ref{62}) of the Bethe-Salpeter
amplitude with respect to the relative times. The unapproximated Bethe-Salpeter equation takes the form
\begin{equation}\Phi=\left[G_{01}G_{02}K^0_{12}+G_{02}G_{03}K^0_{23}+G_{03}G_{01}K^0_{31}+
R\,\right]\Phi\label{83}\end{equation}
and leads to the 3D equation
$$\psi=\int
\!dp_0\,\Lambda^{+++}(1-R)^{-1}\left[G_{01}G_{02}K^0_{12}+G_{02}G_{03}K^0_{23}+G_{03}G_{01}K^0_{31}\right]\psi$$
\begin{equation}=\,\left[\,{\Lambda^{+++}\over
P_0-S}\,(\,V_{12}\,+\,V_{23}\,+\,V_{31}\,)\,\Lambda^{+++}\,+\,\hat R\,\right]\psi\label{84}\end{equation}
with
\begin{equation}\hat R\,=\,\int
\!dp_0\,\Lambda^{+++}\,{R\over
1-R}\,\left[G_{01}G_{02}K^0_{12}+G_{02}G_{03}K^0_{23}+G_{03}G_{01}K^0_{31}\right]\label{85}\end{equation}
In terms of $\,\chi\!=\!(1\!-\!\hat R)\psi:$   
\begin{equation}(P_0-S)\,\chi\,=\,\Lambda^{+++}\,(\,V_{12}\,+\,V_{23}\,+\,V_{31}\,)\,\Lambda^{+++}\,(\,1-\hat
R\,)^{-1}\,\chi.\label{86}\end{equation}
In this last form the correction terms are more symmetric and less total energy dependent.
At the approximation of positive-energy instantaneous two-body kernels, as (\ref{55}), we expect $\,\hat R\,$ (or at least $\,\hat
R\psi\,$) to vanish. This is necessary if we want it to be of higher-order in the general case, but it is not automatic:
although the "approximation" (\ref{82}) led to our basic three-cluster equation, equation (\ref{86}) could turn out to be
a complicated equivalent of it (we met this situation in previous promising attempts of perturbation calculation). For
positive-energy instantaneous two-body kernels, we have
\begin{equation}R\,=\,G_{01}G_{02}K^R_{12}+G_{02}G_{03}K^R_{23}+G_{03}G_{01}K^R_{31}\,\equiv\,R_{12}+R_{23}+R_{31}\label{87}\end{equation}        
\begin{equation}K^R_{12}(p'_{30},p_{30})\,=\,K_{12}\,\left[\,\delta(p'_{30}\!-\!p_{30})\,
\,-\,\Lambda^+_3\,\delta(p'_{30}\!-\!h_3)\,\right].\label{88}\end{equation}
This implies immediately that the terms ending like
$\,R_{12}G_{01}G_{02}K^0_{12}\,$ are zero. For terms ending like
$\,R_{12}G_{02}G_{03}K^0_{23}\,$ we have to examine the action of four possible operators at left:
\begin{equation}\int\!dp_0 (...),\quad R_{23},\quad R_{31},\quad R_{12}.\label{89}\end{equation}
The three first ones lead to contour integrals which give zero. With $\,R_{12}\,$
we must perform the $\,p_{120}\,$ integral and then apply again one of the four operators (\ref{89}) at left, etc...  This
recurrence ends necessarily with one of the first three operators (\ref{89}) which gives zero again.\par          
In the general case our correction terms are thus really higher-order terms. Two aspects remain however to be investigated:\par
-- It seems that our method brings corrections to the two-fermion potentials as well. At the cluster-separated limits, our
two-fermions equations would then be replaced by equivalent equations, unless we could show that these two-fermion correction
terms vanish.\par
-- Our three-fermion correction terms are a priori not hermitian. Furthermore, they are total energy dependent, like the
unperturbed potentials. As explained in the introduction, these two difficulties are not independent.

\section{Conclusions}
 We have written a 3D equation for three fermions (our basic three-cluster equation)  by combining the three two-body potentials obtained by
an exact 3D reduction of the corresponding two-fermion Bethe-Salpeter equations, putting the spectator fermion on the
mass shell. In this way, the cluster-separated limits are still exact, and the   Lorentz invariance / cluster
separability requirement is automatically satisfied, provided no supplementary approximation, like the Born approximation, is made. This equation can
be written  in terms of the two-body potentials, or in terms of two-body transition operators
(Faddeev formalism). The use of positive free-energy projectors in the chosen reductions of the two-fermion Bethe-Salpeter equations   prevents our
3D three-fermion equation from continuum dissolution. The potentials are hermitian and depend only slowly on the total three-fermion energy.  The one
high-mass limits of our "basic three-cluster equation" are approximately exact. The correction of the remaining discrepancy would demand the
introduction of higher-order three-body terms.    
\par  Our combination of cluster separability and Lorentz invariance in the three fermion problem makes explicit use of the fact
that the clusters can only be two-fermion and/or free fermion states. This is not directly adaptable to four and more fermion
systems. In this respect, 3 is still not N.\par
The difference between our equation and Gross' spectator model equations consists in higher-order three-body contributions.
Supplementary investigations would be necessary to decide which approach provides the best approximation to the three-fermion
Bethe-Salpeter equation.\par A lot of work remains to be done in order to precise a possible perturbation calculation program,
correcting the remaining discrepancies with the three-fermion Bethe-Salpeter equation. We succeeded in deriving our 3D equation
from an approximation of the three-fermion Bethe-Salpeter equation, in which the three-body kernel is neglected while the two-body
kernels are approached by positive-energy instantaneous expressions, with the spectator fermion on the mass shell. The correction
terms are thus known at the Bethe-Salpeter level and can be transformed into corrections to the 3D
equation, unfortunately not hermitian a priori.\par 
  There exists an infinity of ways of performing the 3D reduction of the two-fermion Bethe-Salpeter equation. The potentials generated by these
reductions could all by used to build a three-fermion 3D equation, keeping however in mind the continuum dissolution problem. This leaves us a
large freedom to suit phenomenological needs.\par  Our two-body potentials are the sum of an infinity of contributions symbolized by Feynman graphs.
Keeping only  the first one (Born approximation) or a finite number of them renders the Lorentz covariance of the two-fermion clusters only
approximate. A Born approximation preserving the  Lorentz invariance / cluster separability property can be obtained by using  another 3D reduction
based on a covariant second-order two-body propagator of Sazdjian, combined with a covariant substitute of
$\,\Lambda^{++}.$ This leads to a 3D three-cluster equation which is  covariantly Born approximable, but more complicated
\cite{20}.

\end{document}